\documentstyle[amssymb,aps,preprint]{revtex}
\def\be{\begin{equation}}
\def\ee{\end{equation}}
\def\bea{\begin{eqnarray}}
\def\eea{\end{eqnarray}}

\begin{document}
\title{Phase diagram of the CuO$_{3}$ chains in YBa$_{2}$Cu$_{3}$O$_{6+x}$ and PrBa$%
_{2}$Cu$_{3}$O$_{6+x}$ }
\author{R. Franco$^{a}$ and A. A. Aligia$^{b}$}
\address{$^{a}$ Instituto de F{\'{\i }}sica, Universidade Federal Fluminense\\
(UFF), C.P 100.093, Avenida Litor\^{a}nea s/n, 24210-340 Niter\'{o}i, Rio de%
\\
Janeiro, Brasil\\
$^{b}$Comisi\'{o}n Nacional de Energ{\'{\i }}a At\'{o}mica,\\
Centro At\'{o}mico Bariloche and Instituto Balseiro, 8400 S.C. de Bariloche,%
\\
Argentina}
\maketitle

\begin{abstract}
We use a mapping of the multiband Hubbard model for CuO$_{3}$ chains in $R$Ba%
$_{2}$Cu$_{3}$O$_{6+x}$ ($R=$Y or a rare earth) onto a $t-J$ model and the
description of the charge dynamics of the latter in terms of a spinless
model, to study the electronic structure of the chains. We briefly review
results for the optical conductivity and we calculate the quantum phase
diagram of quarter filled chains including Coulomb repulsion up to that
between next-nearest-neighbor Cu atoms $V_{2}$, using the resulting
effective Hamiltonian, mapped onto an $XXZ$ chain, and the method of
crossing of excitation spectra. The method gives accurate results for the
boundaries of the metallic phase in this case. The inclusion of $V_{2}$
greatly enhances the region of metallic behavior of the chains.
\end{abstract}


There is consensus in that the electronic structure of $R$Ba$_{2}$Cu$_{3}$O$%
_{6+x}$ ($R=$Y or a rare earth) can be separated into that of the two CuO$%
_{2}$ planes per unit cell which become superconducting under doping, and
that of the CuO$_{2+x}$ subsystem, in which CuO$_{3}$ chains are formed for
oxygen content $x\geq 0.5$ and low temperatures.\cite{garc} The electronic
structure of the CuO$_{3}$ chains is crucial because it controls the doping
of the superconducting CuO$_{2}$ planes. The dependence of the
superconducting critical temperature $T_{c}$ with annealing,\cite{veal}
combined with Raman measurements \cite{kirch} and persistent
photoconductivity experiments \cite{kudi,nieva} show an intimate relation
between the oxygen ordering in the CuO$_{x}$ planes and $T_{c}$ \cite{com}:
oxygen ordering along chains increases the amount of two-fold and four-fold
coordinated Cu atoms at the expense of three-fold coordinated ones, and
leads to an increase in the hole doping of the superconducting CuO$_{2}$
planes. Detailed calculations of the relation between electronic and atomic
structure in $R$Ba$_{2}$Cu$_{3}$O$_{6+x}$, together with a simple
explanation of the above facts valid in the strong coupling limit were
presented.\cite{garc} These results show the relevance of interatomic
Coulomb interactions. In addition even near the optimum doping ($\sim 1/5$
holes per Cu atom in the planes), the average distance between carriers is
of the order of two lattice parameters of the planes suggesting that
interatomic repulsion at smaller distances are screened only partially.

Several pieces of evidence suggest that the CuO$_{3}$ chains are insulating.
For example PrBa$_{2}$Cu$_{3}$O$_{7}$ is semiconducting,\cite{sode} and the
contribution of the CuO$_{3}$ chains to the optical conductivity $\sigma
(\omega )$ is very similar in this compound \cite{take} and in
superconducting YBa$_{2}$Cu$_{3}$O$_{6+x}$,\cite{schle,rott} displaying a
broad peak near $\omega \sim 0.2$ eV and a slowly falling tail at higher
frequencies. Also, charge modulations observed by scanning tunneling
microscopy (STM) were interpreted in terms of a charge density wave and a
gap in the spectrum of the chains.\cite{edwa}. Finally in the explanation of
Fehrenbacher and Rice of the suppression of superconductivity upon
substituting Y by Pr in YBa$_{2}$Cu$_{3}$O$_{7}$, they propose that the
holes which dope the superconducting CuO$_{2}$ planes in YBa$_{2}$Cu$_{3}$O$%
_{7}$ are displaced towards a hybrid Pr-O state in PrBa$_{2}$Cu$_{3}$O$_{7}$.%
\cite{fehr} This implies a shift in the Fermi level of about 0.25 eV
according to their parameters, while the authors {\em assume} that the hole
occupation of the CuO$_{3}$ chains is 0.5 in both cases. Thus, this
explanation seems to require a gap in the chains to be consistent.

However, all he above data can also be consistently explained assuming
intrinsically {\em metallic} chains cut by $\sim 5\%$ of defects or oxygen
vacancies ($x\sim 0.95$), which is usual in these systems.\cite{gag} The
appropriate multiband model for CuO$_{3}$ chains was mapped numerically into
a $t-J$ model with $t\sim 0.85$ eV and $J\symbol{126}0.2$ eV. With these
parameters, the decrease in the occupation of the chains upon replacing Y by
Pr is only 0.05. Taking into account that the charge dynamics of the model
can be described up to a few percent by a spinless model even for $J/t=0.4$,%
\cite{gag,tohy} charge modulations and the optical conductivity can be
explained.\cite{gag} In particular, the lower energy part of the latter is
given by:

\begin{equation}
\sigma (\omega )=\frac{AB}{\omega ^{2}}\exp (-A/\omega ),  \label{cond}
\end{equation}
\noindent with 
\[
A=-2t\pi \ln (1-c)\sin ~k_{F};~~~~B=\frac{e^{2}}{\hbar }(1-c^{2})t\sin
~k_{F}, 
\]
and $c=1-x$ is the concentration of oxygen vacancies\noindent . The
experimental results were fitted using $A=0.35$ eV.\cite{gag} The resulting
optical conductivity below 0.4 eV is shown in Fig. 1. Note that in spite of
the metallic character of the chains, as a consequence of the oxygen
defects, $\sigma (\omega )$ has a pseudogap at low energies. The experiments
cannot confirm this due to large errors for $\omega <0.1$ eV.\cite
{take,schle,rott} However, recent STM\ studies of the local density of
states detect a pseudogap of about 25 meV and numerous intragap resonances.%
\cite{derro} The latter might be explained by the effect of defects on
superconductivity in the chains induced by proximity,\cite{morr} but also in
principle by eigenstates of long finite metallic chains. Unfortunately, the
local density of states of the one-dimensional $t-J$ model depends also on
the spin wave function and cannot be described solely by spinless fermions.%
\cite{penc,kim}

The natural candidate to open a gap in the effective $t-J$ model for the CuO$%
_{3}$ chains is the nearest-neighbor repulsion $V_{1}$. Keeping the
assumption that the charge dynamics is described by a spinless model, one
expects that a gap opens for $V_{1}>2t\sim 1.7$ eV.\cite{john,zl} If the
Coulomb repulsions were completely unscreened $V_{1}\sim e^{2}/b\cong 3.6$
eV, where $b$ is the lattice parameter along the chains. Recently Seo and
Ogata showed that inclusion of next-nearest-neighbor repulsion $V_{2}$
enhances the range of stability of the metallic phase, calculating the gap
as a function of $V_{2}$.\cite{seo}

We calculate the phase diagram of the spinless model, including $V_{1}$ and $%
V_{2}$ using the method of crossing of excitation levels.\cite{nomu,naka,ih}
Actually, the mapping of the energy of the one-dimensional $t-J$ model into
that of a spinless model is strictly valid only for $J=0$,\cite{penc,zl,seo}
but we expect it to be a very good approximation for $J/t<0.4$.\cite
{gag,tohy} The advantage of the method of level crossings, briefly explained
below, over previous approaches \cite{seo,zhur} is the accuracy that can be
achieved for the phase boundaries. This has been shown for example in its
application to the Hubbard model with correlated hopping \cite{bos,topo} in
comparison with exact results. \cite{afq}

In standard notation, the model is:

\begin{equation}
H=\sum_{i}[t(c_{i+1}^{\dagger }c_{i}+\text{H.c.}%
)+V_{1}n_{i}n_{i+1}+V_{2}n_{i}n_{i+2}],  \label{h1}
\end{equation}
with $n_{j}=c_{j}^{\dagger }c_{j}$. Using a Jordan-Wigner transformation $%
S_{j}^{+}=c_{j}^{\dagger }\exp (i\pi \sum_{l<j}n_{l})$, $%
S_{j}^{-}=(S_{j}^{+})^{\dagger }$, $S_{j}^{z}=n_{j}-1$, the model can be
mapped into an $XXZ$ model with next-nearest-neighbor antiferromagnetic
Ising interaction:

\begin{equation}
H=\sum_{i}[J_{1}(S_{i}^{x}S_{i+1}^{x}+S_{i}^{y}S_{i+1}^{y})+\Delta
_{1}S_{i}^{z}S_{i+1}^{z}+\Delta _{2}S_{i}^{z}S_{i+2}^{z}],  \label{h2}
\end{equation}
where $S_{i}^{\beta }$ is the $\beta $ component of the spin-1/2 operator at
site $i$, $J_{1}=2t$ and $\Delta _{j}=V_{j}$.\cite{zl,seo}

A successful approach to describe the qualitative properties of
one-dimensional strongly correlated systems is bosonization followed by a
renormalization group procedure. This procedure usually terminates at a
fixed point, which determines the properties of the system for the initial
parameters given. A phase transition occurs when the flow goes towards a
different fixed point. Since the renormalization group is a weak coupling
approach, the phase boundaries are not given accurately by the method for
large interactions. The basic idea of the method of level crossings is to
combine numerical calculations of excitation levels with basic knowledge on
the properties of these fixed points. The more interesting phase transitions
involve one fixed point which is scale invariant. This is for example the
case of the $XXZ$ model with next-nearest-neighbor interactions studied by
Nomura and Okamoto.\cite{nomu} The spin fluid phase of Eq. (\ref{h2}) (which
corresponds to the metallic phase of Eq. (\ref{h1})), like that of an
ordinary Heisenberg model is characterized by a scale invariant fixed point.%
\cite{nomu} Then, using conformal field theory one can relate the excitation
energy which corresponds to some operator $A_{i}$ at site $i$ (for example a
spin flip $S_{i}^{+}$, $S_{i}^{-}$), to the dependence of the correlation
functions of this operator with distance $d$, for large $d$: 
\begin{equation}
E_{A}(L)-E_{g}(L)=\frac{2\pi vx_{A}}{L},\text{ }\left\langle
A_{i+d}A_{i}\right\rangle \backsim \frac{1}{d^{2x_{A}}}.  \label{ee}
\end{equation}
Here $L$ is the length of the system, $v$ the spin-wave velocity, $E_{g}(L)$
the ground state energy, $E_{A}(L)$ the lowest energy in the adequate
symmetry sector (connected to the ground state by $A_{i}$) and $x_{A}$ the
critical dimension for the excitation $A$. Since the dominant correlations
at large distances determine the nature of the thermodynamic phase, a phase
transition is determined by the crossing of excited levels for different
symmetry sectors.

In the present problem, the relevant quantum numbers which determine the
symmetry sector are total wave vector $K$, total spin projection $S^{z}$,
parity under inversion $P$ and parity under time reversal $T$ . We have
restricted our calculations to number of sites $L$ multiple of four to avoid
frustration of the phase which we call AFII (see below). For these sizes,
the quantum numbers of the ground state are always the same in the region of
parameters studied. They are listed in Table I, together with the quantum
numbers of the first excited state of each phase. Our main interest are the
boundaries of the spin fluid phase of the spin model Eq. (\ref{h2}) which
corresponds to the metallic phase of Eq. (\ref{h1}). With increasing $\Delta
_{1}$ ($\Delta _{2}$) there is a continuous transition to an insulating Neel
ordered (dimerized) phase.\cite{nomu} The Neel ordered phase, which we call
antiferromagnetic I (AFI) for maximum order parameter has a spin ordering $%
\uparrow \downarrow \uparrow \downarrow ...$ and corresponds to a charge
ordering 1010... in the original model Eq. (\ref{h1}). The dimer phase has a
gap which is exponentially small near the metallic phase \cite{nomu}. This
renders it very difficult to detect the transition with alternative
numerical methods.\cite{topo} The transitions between any two of these three
phases were determined accurately from the corresponding crossing of excited
levels (see Table I). In addition, with increasing $\Delta _{2}$, we expect
a transition from the dimer phase to an AFII a phase with long range order $%
\uparrow \uparrow \downarrow \downarrow ...$ (corresponding to charge
ordering 1100...). This transition cannot be detected by crossing of first
excited states. Since it involves two insulating phases, it is not described
by a scale invariant theory and is also beyond our scope. For the sake of
completeness we have drawn a tentative dimer-AFII boundary using the rough
criterium that the system is in the AFII phase when the ground state
correlation function (calculated deriving the energy using Hellmann-Feynman
theorem) $\langle S_{i}^{z}S_{i+2}^{z}\rangle <-1/8$. For the other
transitions, we have calculated the transition points in systems with $L=12$%
, 16 and 20 sites. According to field theory predictions for large enough $L$%
, these points plotted as a function of $1/L^{2}$ should lie on a straight
line.\cite{nomu,note} We have verified that this is the case for the three
transitions with high accuracy. The linear fit provided the transition point
extrapolated to $1/L^{2}\rightarrow 0$, and is error. The error is below 1\%
in all cases, confirming the validity of the method in the present case.

The resulting phase diagram is shown in Fig. 2.  For $V_{2}=\Delta _{2}=0$,
the known exact results \cite{john,zl} are reproduced: there is a transition
from the spin fluid (metallic) phase to the AFI (charge density wave) phase
at $\Delta _{1}=J_{1}$ ($V_{1}=2t$). Another known limit is the classical
one $J_{1}\rightarrow 0$ ($t\rightarrow 0$), for which there is a transition
between both AF phases at $\Delta _{2}=\Delta _{1}/2$ ($V_{2}=V_{1}/2$). Our
results are consistent with this limit. However, there is a strip of width $%
\sim J_{1}=2t$ of a dimer phase between both AF (charge ordered insulating)
phases. This is reminiscent of the physics of the ionic Hubbard model, for
which a strip of a dimer phase of width $\sim 0.6t$ in the strong coupling
limit, separates the band insulating and the Mott insulating phases \cite{ih}
dye to the charge fluctuations that still remain in the strong coupling
limit.

In qualitative agreement with previous calculations,\cite{seo} we obtain
that the addition of $V_{2}$ greatly enhances the range of stability of the
metallic phase of the CuO$_{3}$ chains in $R$Ba$_{2}$Cu$_{3}$O$_{6+x}$. For
unscreened interactions $V_{1}=2V_{2}\cong 3.6$ eV. Using $J_{1}=2t=1.7$ eV,
one can see from the phase diagram, that the system falls in the metallic
phase even in this extreme case. Instead, if $V_{2}$ were neglected the
chains would be in an insulating charge ordered state for the same $t$ and $%
V_{1}$.

A numerical mapping of the appropriate multiband Hubbard model for the
chains to a $t-J$ model indicates that $J/t<1/4$ \cite{gag}. It is
reasonable to expect that turning $J$ to zero does not change sustantially
the phase diagram. For $J=0$, the mapping to the spinless model is exact and
our results lead to the conclusion that the CuO$_{3}$ chains in $R$Ba$_{2}$Cu%
$_{3}$O$_{6+x}$ ($R=$Y or a rare earth) are intrinsically metallic. Observed
charge modulations are likely due to Friedel oscillations induced by
defects, like O vacancies. These defects or superconductivity induced by the
CuO$_{2}$ planes can also lead to the observed pseudogap behavior.

This work was sponsored by PICTs 03-06343 of ANPCyT, Argentina. R. Franco is
grateful to the National Research Council (CNPq), Brasil, by their financial
support. A.A. Aligia is partially supported by CONICET, Argentina.

\section*{Table I}

\begin{tabular}{|c|c|c|c|c|}
\hline
& $K$ & $S^{z}$ & $P$ & $T$ \\ \hline
{\ \ ground state} & {\ 0} & {\ 0} & {\ 1} & {\ 1} \\ \hline
{\ exc. spin fluid} & {\ }$\pi $ & {\ }$\pm ${1} & {\ -1} & {\ -} \\ \hline
{\ exc. AFI} & {\ }${\pi }$ & {\ 0} & {\ -1} & {\ -1} \\ \hline
{\ exc. dimer, AFII} & {\ }${\pi }$ & {\ 0} & {\ 1} & {\ 1} \\ \hline
\end{tabular}

Quantum numbers of the ground state and the first excited state of the
different phases for $L$ multiple of four.

\section*{Figure captions}

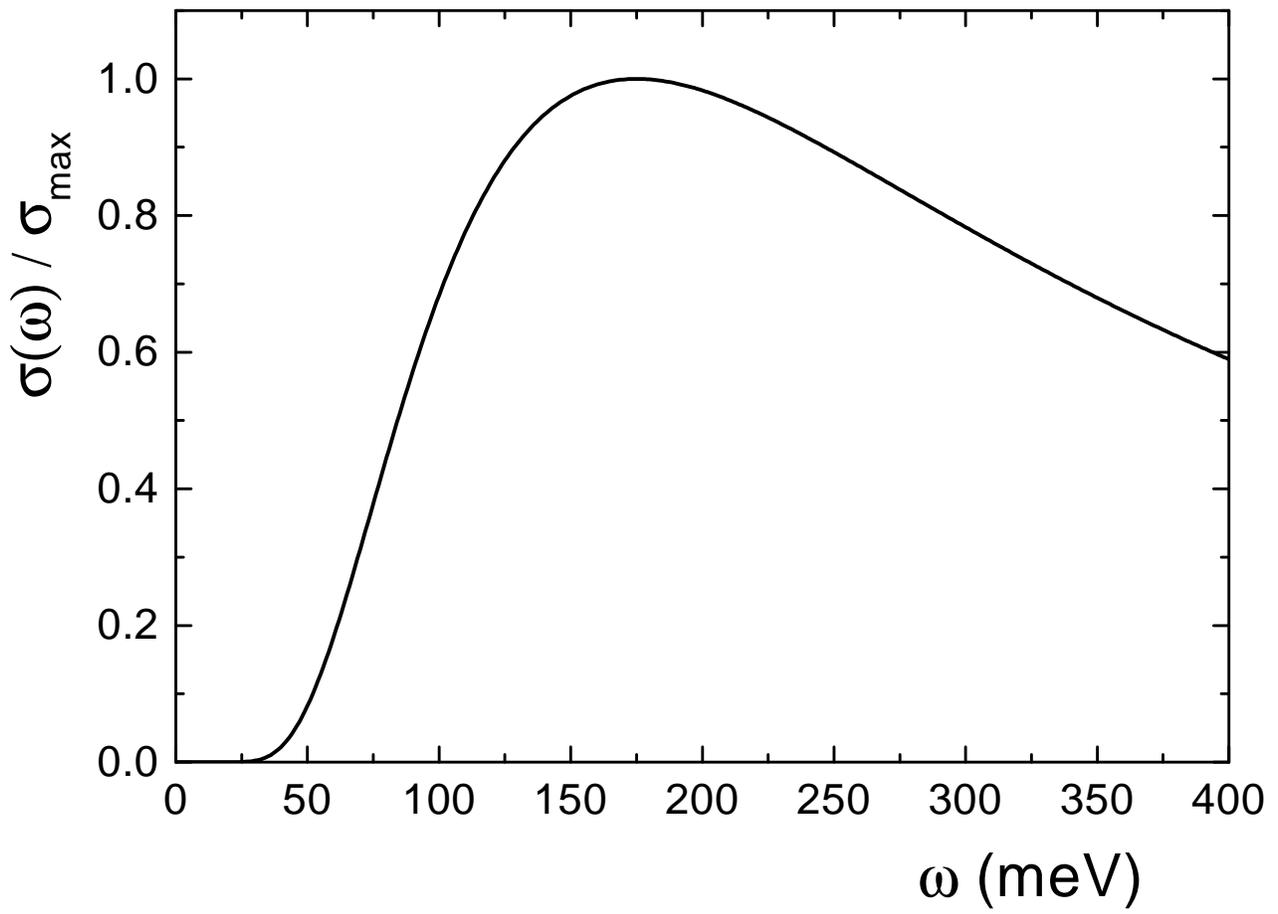
{\bf Fig.1. } Low energy part of the optical conductivity of CuO$_{3}$
chains for $A=0.35$ eV (see Eq. (\ref{cond})).

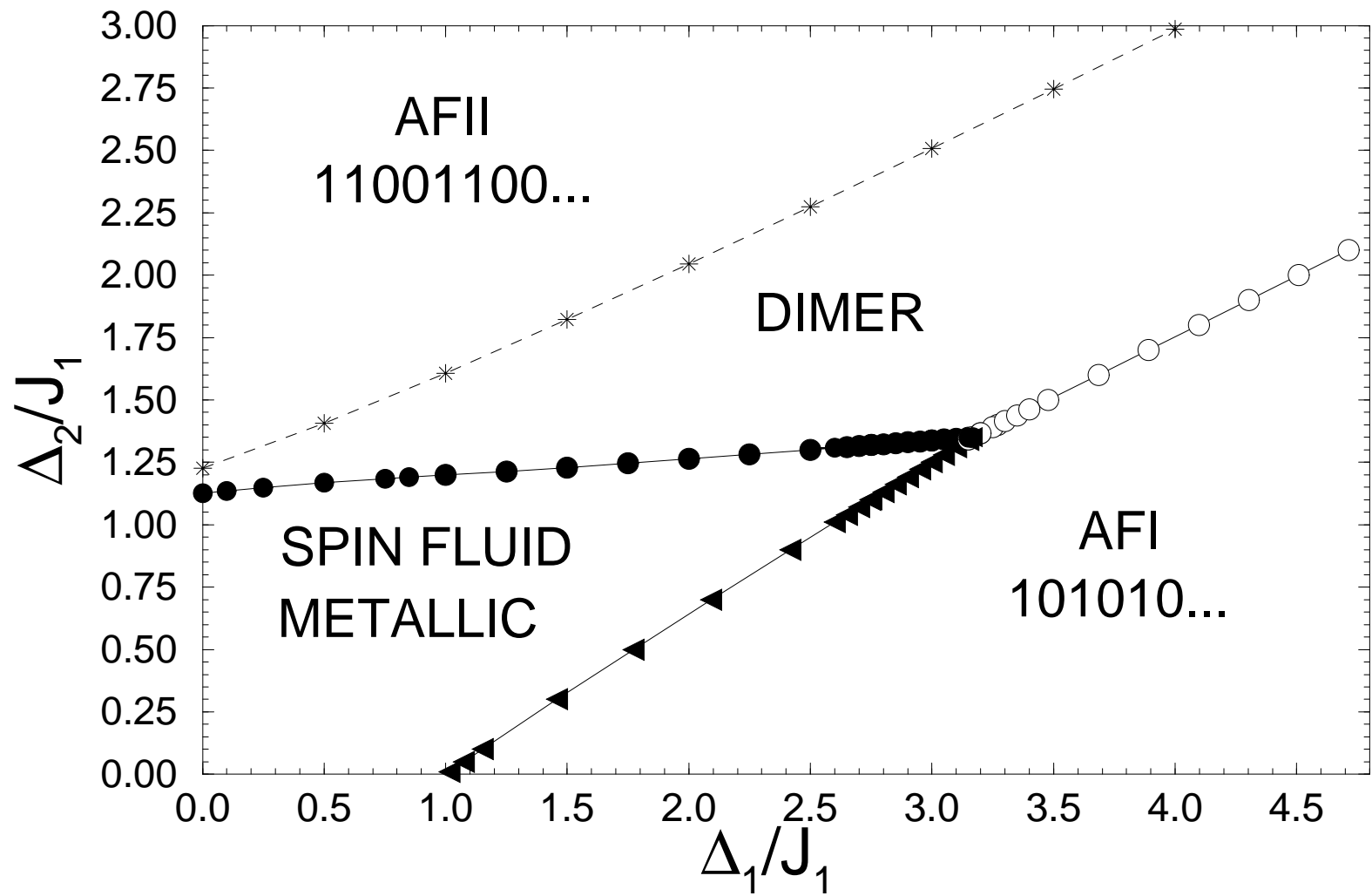
{\bf Fig.2. } Phase diagram of the effective model for CuO$_{3}$ chains (Eq.
(\ref{h1}) or (\ref{h2})) as a function of $\Delta _{1}/J_{1}=V_{1}/2t$ and $%
\Delta _{2}/J_{1}=V_{2}/2t.$


\begin{references}
\bibitem{note}  Actually, the $1/L^{2}$ scaling is not expected to work for
the transition between gapped phases, like AFI and dimer, unless the gap is
very small at the transition.
\end{references}
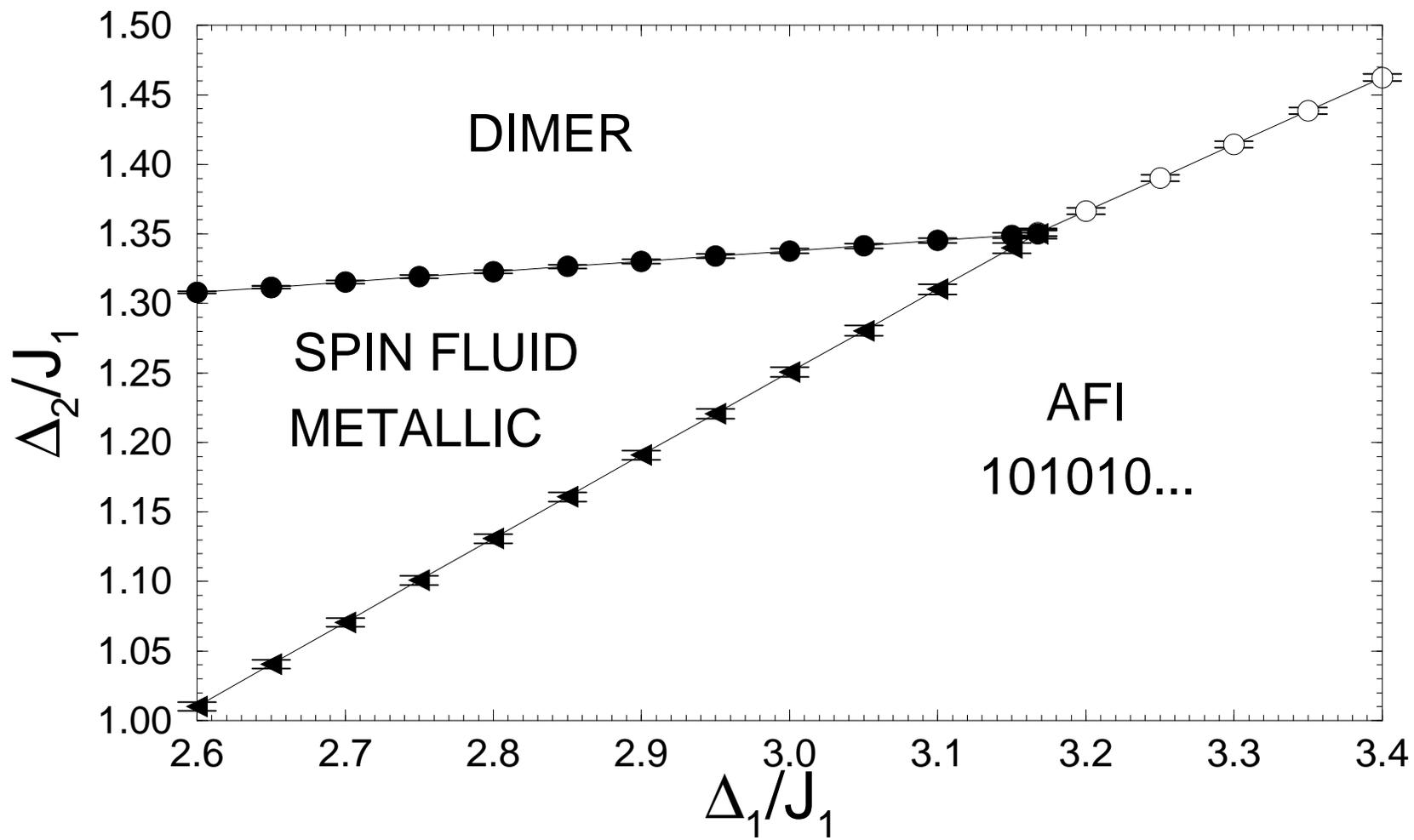
\end{document}